\documentclass{article}
\usepackage{graphicx}
\usepackage{epsfig}
\usepackage{amsfonts}
 \usepackage{amssymb}
\usepackage{xcolor}
\usepackage{mathtools}
\usepackage[utf8]{inputenc}
\usepackage[english]{babel}
\usepackage{amsmath,amsfonts,amssymb,amsthm}
\date{\today}

\usepackage{amsmath}
\usepackage{amssymb}

\newcommand{\diff}{\mathrm{d}}

\usepackage{xcolor}
\newcommand{\be}{\begin{equation}}
\newcommand{\ee}{\end{equation}}

\begin{document}

 \title{Boson and neutron stars with increased density} 

\author{
{\large Yves Brihaye}$^{(1)}$, {\large Ludovic Ducobu}$^{(2)}$  and 
{\large Betti Hartmann}$^{(3),(4)}$
\\ 
\\
$^{(1)}${\small D\'ep. Physique de l'Univers, Champs et Gravitation, Universit\'e de
Mons, Mons, Belgium}\\
$^{(2)}${\small D\'ep. Physique Nucl\'eaire et Subnucl\'eaire, Universit\'e de
Mons, Mons, Belgium}\\
$^{(3)}${\small Instituto de F\'isica de S\~ao Carlos, Universidade de S\~ao Paulo, Brazil }\\
$^{(4)}${\small Department of Theoretical Physics, University of the Basque Country, Bilbao, Spain} }

\maketitle 
\begin{abstract} 
We discuss boson stars and neutron stars, respectively, in a scalar-tensor gravity model with an explicitly time-dependent real scalar field. 
While the boson stars in our model -- in contrast to the neutron stars -- do not possess a hard core, we find that the qualitative effects of the scalarization are similar in both cases~: the presence of the gravity scalar
allows both type of stars to exist for larger central density as well as larger mass at given radius than their General Relativity counterparts. In particular, we find new types of scalarized neutron stars which have radii very close to the corresponding Schwarzschild radius and hence are comparable in density to black holes. 
\end{abstract}

\section{Introduction}
With increased interest in astrophysical objects and, in particular, their gravitational properties,
compact objects have come to the focus of theoretical research again. These objects are normally defined to have strong gravitational fields and as such are a good testing ground for
the momentarily accepted best model of the gravitational interaction -- General Relativity (GR) --
as well as extensions thereof and even alternative gravity models. Compact objects come in two varieties:
either they are star-like with a globally regular space-time or they possess a physical
singularity shielded from observation by an event horizons. The former 
are neutron stars and  boson stars, respectively, the latter black holes. 
While neutron stars and black holes are known to exist and can now be studied with unprecedented
precision, boson stars \cite{bosonstars}  are hypothetical objects made principally of scalar bosonic particles. 
Evaluating and testing gravity theories is also vital in order to understand two of the great puzzles
of current day physics~: the nature of dark matter and dark energy. While dark matter is
understood to be some kind of matter that interacts only gravitationally and probably has its origin in physics beyond the Standard Model of Particle physics, the nature of dark energy remains elusive. 
Consequently, suggestions for a modification of GR have been made on the ground of
so-called ``scalar-tensor'' gravity models \cite{Nicolis:2008in,Deffayet:2011gz,Deffayet:2013lga},
an idea that relates back to Horndeski \cite{horndeski}. Classes of scalar-tensor gravity models
have then been studied thoroughly and a  classification, named ``Fab Four'', was achieved \cite{CCPS1,CCPS2}.
In this paper, we are interested in a particular model dubbed ``John'' in this exact classification.
As has been shown in \cite{babichev}, static, spherically symmetric black holes can carry
scalar hair in this model if the scalar field is explicitly (and linearly) time-dependent.  In particular, the Noether current associated to the shift symmetry of the Galileon-type gravity scalar does not
diverge on the horizon in this model. In \cite{cdr2015}, neutron stars have been studied
for a specific polytropic equation of state and it has been claimed that the astrophysical
objects resulting from the model are viable and not in conflict with constraints from observations.
Here, we revisit these results  and compare them with those related to another EOS used in
\cite{lemos2004}. We find that the solutions obtained with the EOS of \cite{lemos2004} 
(a) are in perfect agreement with results obtained in \cite{lemos2004} and 
(b) only this EOS leads to  neutron stars  possessing the proper
mass-radius relation. While neutron stars are matched to the Schwarzschild solution at the exterior
radius, we also discuss boson stars in this paper that reach the Schwarzschild solution
only asymptotically and hence do not possess a ``hard core''.

Our paper is organized as follows: in Section \ref{section:model} we discuss the scalar-tensor
gravity model coupled to an appropriate energy-momentum content. In Section \ref{section:BS}, we present our results for boson stars, while Section \ref{section:NS} contains our findings for neutron stars. We summarize and conclude in Section \ref{section:summary}.

\section{The model}
\label{section:model}
In this paper, we present our results for a scalar-tensor gravity model of
Horndeski type coupled minimally to an appropriate matter content
with Lagrangian density ${\cal L}_{\rm matter}$. The action reads~:
\begin{equation}
{\cal S} = \int \left(\kappa {\cal R}+ \frac{\eta}{2} G^{\mu\nu} \nabla_\mu \phi \nabla_\nu\phi + {\cal L}_{\rm matter}\right) \sqrt{- g}\diff^4 x \ ,
 \end{equation}
 where $\kappa=(8\pi G)^{-1}$.  This action contains the standard Einstein-Hilbert term as well
 as a non-miminal coupling term -- first discussed in \cite{CCPS1,CCPS2} -- that couples
 a gravity scalar $\phi$ to the Einstein tensor $G_{\mu\nu}$ via a coupling constant $\eta$.
 For $\eta=0$, we recover standarad General Relativity (GR). 
 
 In the following, we will assume the matter content of the model to be that of (a) a complex valued scalar field
 and (b) a perfect fluid with a given equation of state, respectively. In the latter case, the model 
 has solutions in the form of neutron stars, while the complex scalar field in curved space-time
 describes boson stars. The gravity equations then read
 \begin{equation}
 \label{eq:gravity}
 \kappa G_{\mu\nu}  + \eta \left(\partial_{\alpha}\phi \partial^{\alpha} \phi G_{\mu\nu} -\frac{1}{2} \epsilon_{\mu\alpha\sigma\rho} R^{\sigma
 \rho \gamma\delta} \epsilon_{\nu\beta\gamma\delta} \nabla^{\alpha} \phi \nabla^{\beta} \phi + g_{\mu\alpha} \delta^{\alpha\rho\sigma}_{\nu\gamma\delta} \nabla^{\gamma} \nabla_{\rho} \phi \nabla^{\delta} \nabla_{\sigma}\phi\right)=T_{\mu\nu}  \ ,
 \end{equation}
 where $T_{\mu\nu}$ denotes the energy-monentum tensor of the matter content.
The model has a shift symmetry $\phi\rightarrow \phi + c$, where $c$ is a constant, which leads to the existence of a locally conserved Noether current
 \begin{equation}
 \label{eq:Noether_scalar}
 J^{\mu}=- \eta G^{\mu\nu} \nabla_{\nu} \phi \ \ , \ \  \nabla_{\mu} J^{\mu}= 0  \ .
 \end{equation}
 In the following, we will assume a spherically symmetric Ansatz for our solutions \cite{babichev}
\begin{equation}
\diff s^2 = - b(r) \diff t^2 + \frac{\diff r^2}{f(r)} + r^2 \left(\diff\theta^2 + \sin^2 \theta \diff\varphi^2\right) \ , \
\phi(t,r) = q t + F(r)  \ ,
 \end{equation}
 i.e. the tensor part is static, while the gravity scalar has an explicit time-dependence.  The non-vanishing components of the Noether current (\ref{eq:Noether_scalar}) then read
 \begin{equation}
 J^t =  \eta q  \frac{  f'r  + f  -  1}{r^2 b}\ \ , \ \  J^r=\eta \phi' 
 \frac{f\left(-b'r f - bf + b\right)}{ r^2 b} \ \ , \ \ 
 \end{equation}
 where the prime now and in the following denotes the derivative with respect to $r$.
The norm of the Noether current is 
\begin{equation}
\label{eq:norm}
J_{\mu}J^{\mu}=\eta^2 \left[-q^2 \frac{ (f'r  + f - 1)^2}{r^4 b} +
\phi'^2 \frac{f\left(b'r f + bf - b\right)^2}{r^4 b^2} \right] \ .
\end{equation} 
Since $f(r \ll 1)\sim 1+ f_2 r^2$ and $b(r\ll 1)\sim 1+ b_2 r^2$ with $f_2$, $b_2$ constants (see below for explicit expressions), the norm of the Noether current is finite for all $r\in[0:\infty)$. 
 
We want to consider a non-vanishing energy-momentum tensor that sources the
tensor and scalar gravity fields.  In the following, we will choose the energy-momentum tensor to be of the form
\begin{equation}
T_{\mu}^{\nu}={\rm diag}(-\rho,P_r,P_t,P_t) \ ,
\end{equation}
where $\rho$ is the energy-density, while $P_r$ and $P_t$ are the radial and tangential
pressures, respectively.  The gravity equations are then a set of coupled, non-linear ordinary differential equations that have to be solved numerically. However, we can simplify the analysis by noting that the equation for the gravity scalar $\phi$, which comes from the $rr$-component of (\ref{eq:gravity}), can be solved algebraically in terms of the other functions~:
\begin{equation}
\label{eq:phip2} 
                      \eta(\phi')^2 = \frac{2 r^2}{ f} P_r + \frac{1-f}{b f} \eta q^2 \ \  .
\end{equation}
This allows the elimination of $\phi$ from the remaining equations and we are left with the equations for the metric functions which read~:  
\begin{equation} 
\label{eq:metric}
           {\cal F}_1 f' + {\cal F}_2 = 0 \ \ ,  \ \  \frac{b'}{b}=\frac{1-f}{fr}
\end{equation}
with
\begin{equation}           
 {\cal F}_1 =4\kappa br +  2b r^3 P_r - 3 \eta q^2 r f \ \ 
\end{equation}
and
 \begin{equation}            
          {\cal  F}_2 =   3 \eta q^2 f(1-f) + 
         2 b\left[\rho r^2 (f+1) + 2 f r^2 P_r + 4 f r^2 P_t+ 2\kappa(f-1)\right] \ \ .
\end{equation}
Note that the second equation in (\ref{eq:metric}) ensures that the Noether current
$J^{\mu}$ is covariantly conserved, i.e. $\nabla_{\mu} J^{\mu}=0$ and is, in fact, the
$rt$-component of the Einstein equation. 

Star-like astrophysical objects are typically characterized in terms of their mass-radius relation. The gravitational mass $M_{G}$ of this solution is given in terms of the asymptotic behaviour of the metric function $f(r)$~:
\begin{equation}
                  f(r) \xrightarrow[r\rightarrow\infty]{ }  1 - \frac{M_{G}}{4\pi \kappa r} + {\cal O}(r^{-2})  \ ,
\end{equation}
while the radius will be defined differently in the case of boson stars and neutron stars, see below.
Since asymptotically, the metric function $b(r)$ becomes equal to $f(r)$ and we assume in the
following that either the pressure $P_r$ tends exponentially to zero asymptotically
(in the case of boson stars) or is strictly zero (in the case of neutron stars), we observe that
the mass $M_G$ can also be read off from the behaviour of the gravity scalar at infinity. Using
(\ref{eq:phip2}) we find that 
\begin{equation}
(\phi')^2 \xrightarrow[r\rightarrow\infty]{ } \frac{M_G}{4\pi \kappa r} q^2  \ .
\end{equation}
In other words: $M_G q^2/(4\pi \kappa)$ constitutes the ``charge'' associated to the scalar field
$(\phi')^2$. 
 
 \section{Boson stars}
 \label{section:BS}
  In the case of boson stars, the energy-momentum content is that of a complex valued
 scalar field $\Psi$, which -- in contrast to the neutron star model discussed in Section \ref{section:NS} -- is not of perfect fluid type. The energy-momentum tensor reads~:
 \begin{equation}
 T_{\mu\nu} = -g_{\mu\nu}\left[\frac{1}{2}g^{\alpha\beta} \left(\partial_{\alpha} \Psi^*
 \partial_{\beta} \Psi + \partial_{\beta} \Psi^* \partial_{\alpha} \Psi\right) + m^2 \Psi\Psi^*\right]
 +\partial_{\mu} \Psi^* \partial_{\nu} \Psi + \partial_{\nu} \Psi^*\partial_{\mu} \Psi  \ ,
 \end{equation}
 where $m$ denotes the scalar boson mass. 
 This model contains an additional conserved Noether current due to the internal global U(1)
symmetry $\Psi \rightarrow \exp(i\chi)\psi$, where $\chi$ is a  constant. This reads
\begin{equation}
j^{\mu} = -\frac{i}{2}\left(\Psi^*\nabla^{\mu} \Psi - \Psi \nabla^{\mu} \Psi^*\right)  \ \ , \ \
\nabla_{\mu} j^{\mu}=0  \ .
\end{equation}  
 With the standard spherically symmetric Ansatz for boson stars
\begin{equation}
\Psi(r,t)=\exp(i\omega t) H(r) \ ,
\end{equation}
where $\omega > 0$ is a constant, the non-vanishing components of the energy-momentum tensor read
\begin{eqnarray}
\label{eq:density_pressure_BS}
\rho = f (H')^2 & + & \left(m^2 + \frac{\omega^2}{b}\right) H^2  \ , \  \nonumber \\
 P_r  =   f (H')^2 - \left(m^2 - \frac{\omega^2}{b}\right) H^2 \  & , & \  
P_t  =   -f (H')^2 - \left(m^2 - \frac{\omega^2}{b}\right) H^2 \ . \ 
\end{eqnarray}
The locally conserved current and associated globally conserved Noether charge are~:  
\begin{equation}
j^t = - \frac{\omega H^2}{b} \  \  \ , \ \ \   Q= - \int \diff^3 x \ \sqrt{-g} \ j^t =
4\pi \omega \int \diff r \ r^2 \ \frac{H^2}{\sqrt{bf}}
\end{equation}
Note that in the model with ungauged U(1) symmetry, the Noether charge $Q$ is frequently interpreted as the number of bosonic particles of mass $m$ that make up the boson star.
Finally the field equation for $\Psi$ reads~:
 \begin{equation}  
 \label{eq:psi}
    H'' + \frac{1}{2}\left( \frac{4}{r} + \frac{f'}{f} + \frac{b'}{b} \right ) H' + \frac{1}{f} \left(\frac{\omega^2}{b} - m^2\right) H = 0   \ .
\end{equation}
The asymptotic behaviour of $H(r)$ that can be read of from (\ref{eq:psi}) is~:
\begin{equation}
\label{eq:h_infty}
H(r)\xrightarrow[r\rightarrow\infty]{ }  \frac{1}{r}\exp\left(-\sqrt{m^2 -\omega^2} r\right) \ ,
\end{equation}
i.e. although the scalar field making up the boson star decays fast, the star does not have a ``hard surface'' like the neutron star discussed below. Rather, its energy density $\rho$ and pressures $P_r$ and $P_t$, respectively, tend to zero only asymptotically. 
We can, however, use an estimate of the radius $R$ of the boson star which is given as follows~:
\begin{equation}
\label{eq:radius}
               \langle R \rangle = \frac{1}{Q} \int  \ \diff^3 x \sqrt{-g} \ r \  j^t  \ = \frac{4\pi \omega}{Q} \int \diff r \ r^3 \frac{H^2}{\sqrt{bf}} \ .
\end{equation}

The equations (\ref{eq:metric}) and (\ref{eq:psi}) have to be solved with boundary conditions that guarantee the regularity of the solution at the origin and its finiteness of energy. The appropriate conditions read : 
\begin{equation}
 b'(0) = 0  \ \ , \ \ H'(0) = 0 \ \ , \ \ b(\infty)=1 \ \ , \ \ H(\infty) = 0
\end{equation}
where the constant $H(0)\equiv H_0$ is an {\it a priori} free parameter that determines the value of $\omega$ as well as the central density of the boson star, see (\ref{eq:density_pressure_BS}), via $\rho(0)=(m^2+\omega^2/b(0))H_0^2$.
As is well known from boson stars in GR, the parameter $H(0)$ can be increased arbitrarily such that a succession of branches of boson stars exist that end only for $H(0)\rightarrow\infty$ and
$b(0)\rightarrow 0$ in this limit. This will be different for the scalar-tensor boson stars studied here. 
 The expansion of the fields around the origin already gives hints that this should be the case. We find~:
\begin{equation}
\label{eq:taylor}
                     b(r) = b_0 \left[1 + \frac{4 H_0^2(2\omega^2 - b_0 m^2)}{3(4\kappa b_0
                      - 3 \eta q^2)} r^2 + {\cal O}(r^4)\right] \ \ , \ \
                     f(r) = f_0\left[ 1 + \frac{1}{6} \left(m^2 - \frac{\omega^2}{b_0}\right) r^2  + {\cal O}(r^4)\right]   \ ,
\end{equation}
where $b_0=b(0)$ and $f_0=f(0)$. This implies that we have to require $4\kappa b_0 - 3 \eta q^2 \neq 0$. 
As we will demonstrate in the following, this condition is crucial in the limitation of the domain of existence of the solutions for $\eta > 0$. Note that for $\eta < 0$ another limitation exists, related to the requirement of positivity of  the right hand side of (\ref{eq:phip2}). 
\\
The system of equations is unchanged under the following rescalings
\begin{equation}
\label{eq:scalings}
r\rightarrow \frac{r}{m} \ \ , \ \ \omega\rightarrow m\omega  \ \ , \ \  H\rightarrow \sqrt{\kappa} H \ \ , \ \ 
\eta \rightarrow \kappa\eta \ \ , \ \ 
\phi \rightarrow \frac{\phi}{m}  \ ,
\end{equation}
which rescales  the radius, mass and Noether charge  of the boson star 
as follows~:
\begin{equation}
\label{eq:scaling_BS}
 \langle R \rangle \rightarrow \frac{\langle R \rangle }{m} \ \ , \ \  M_G \rightarrow 
\frac{M_G}{m}  \  \ , \ \  Q \rightarrow \frac{\kappa}{m^2} Q \ . 
\end{equation}
In the following we will choose $\kappa=1$, $m=1$, $\eta=\pm 1$ without loss of generality.

\subsection{Numerical results} 
We have solved the equations numerically using a collocation method for boundary-value differential equations using damped Newton-Raphson iterations \cite{COLSYS}.
The relative errors of the solutions are on the order of $10^{-6}-10^{-10}$.
The constants to be varied are the combination $\eta q^2$ as well as $\omega$ (or equivalently $H(0)$).  From (\ref{eq:h_infty}) we know that with the rescalings (\ref{eq:scalings}) the angular frequency is restricted by:  $\omega^2 \leq 1$.

\begin{figure}[ht!]
\includegraphics[width=9cm]{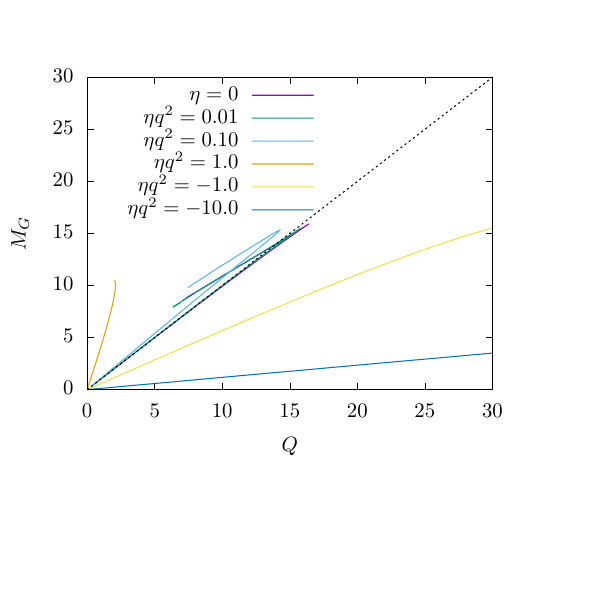}
\includegraphics[width=9cm]{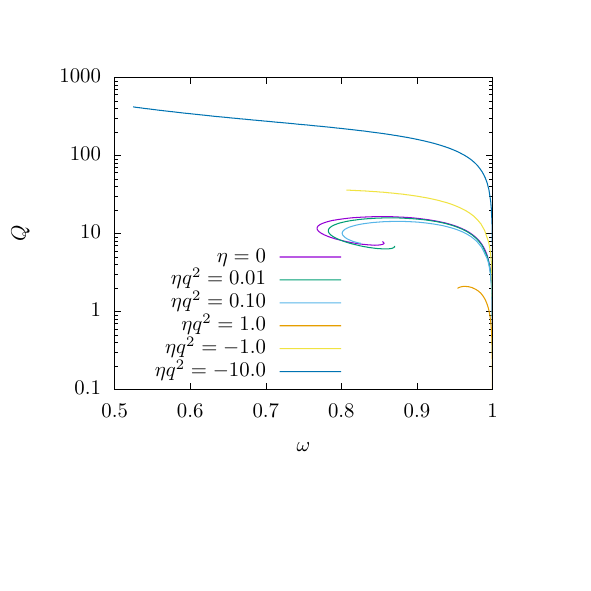}
\vspace{-2.5cm}
\begin{center}
\caption{We show the gravitational mass $M_G$ as function of the Noether charge $Q$ (left) as well as the Noether charge $Q$ as function of $\omega$ (right) for GR boson stars ($\eta=0$) and boson stars with time-dependent
scalar hair for several values of $\eta q^2$. 
\label{fig:M_Q_omega}}
\end{center}
\end{figure}

\begin{table}
\begin{center}
    \begin{tabular}{|c|c|c|c|c|c|c|c|}
    \hline
    $\eta q^2$ &  $M_{G, {\rm max}}$   &  $Q_{\rm max}$  &   
    $\langle R\rangle^*$   & $\omega^*$ & $\rho(0)^*$  & $P(0)^*$  \\    \hline \hline
    $0$ &   $15.91 $   &  $16.40 $  & $3.10 $ & $0.85  $ &  $0.19$ &  $0.04 $\\ \hline
     $0.01$ &   $ 15.50 $   &  $15.81$  & $3.22$  & $0.86$ &  $0.15$ & $0.03$ \\ \hline
     $0.1$ &   $15.32 $   &  $ 14.27$  & $3.54$ & $0.87$ &  $0.12$ & $0.02$   \\ \hline
     $1.0$ &   $10.46 $   &  $2.10 $  & $6.44$  & $0.96$ &  $3 \cdot 10^{-3}$ & $2\cdot 10^{-4}$ \\ \hline
     $-1.0$ &   $16.68$   &  $ 35.93$  & $2.84  $   & $0.83$ &  $0.47$ & $0.11$\\ \hline
     $-10.0$ &   $ 17.90$    &  $416.81$  & $ 2.50  $  & $0.79$ &  $4.13$ & $1.00$ \\ 
        \hline
    \end{tabular}
\end{center}
\caption{We give the maximal values of the mass $M_{G,{\rm max}}$ as well as the maximal value of the Noether charge $Q_{\rm max}$ for different values of $\eta q^2$. Also given is the mean radius $\langle R\rangle^*$, the angular frequency $\omega^*$, the central density $\rho(0)^*$
as well as the central pressure $P_r(0)^*=P_t(0)^*\equiv P(0)^*$ at the maximal value of the mass, i.e. at $M_{G,{\rm max}}$.}
\label{table1}
\end{table}

In Fig. \ref{fig:M_Q_omega} we show the relation between Noether charge $Q$ and
gravitational mass $M_G$ (left) and the dependence of the Noether charge on $\omega$ (right), respectively, for several values of $\eta q^2$ including the GR case $\eta=0$.
While for $\eta=0$, we can increase the value of $H(0)$ arbitrarily, this is no longer the case
in the scalar-tensor gravity model studied here. For $\eta q^2 >0$, the curves shown in
Fig. \ref{fig:M_Q_omega} are limited by the requirement discussed above which, with our choice of constant,  reads~: $4 b_0 - 3 \eta q^2 > 0$. We find that the branches of solutions stop at $4b_0 - 3 \eta q^2 = 0$. For the GR case and $H(0)\rightarrow \infty$ the value of the metric function $b(r)$ at $r=0$, $b_0$, tends to zero.
This is obviously no longer true and hence boson stars with time-dependent scalar hair
are limited in their central density of the star. For $\eta q^2$ sufficiently large, see the curves
for $\eta q^2=1.0$, this also leads to the observation that the Noether charge $Q$ is strongly limited and much smaller than in the GR case. On the other hand, the mass $M_G$ is of the same order of magnitude. Hence, scalar-tensor boson stars with time-dependent scalar fields
and $\eta q^2 > 0$ are comparable in mass, but  consist of an order of magnitude smaller number of  scalar bosonic particles as compared to their GR counterparts. Moreover, their central
density $\rho(0)$ and central pressure $P_r(0)=P_t(0)\equiv P(0)$ is comparable to the GR case, see Table \ref{table1} as long as $\eta q^2$ is not too large. For $\eta q^2=1.0$, we find that
both the central density as well as the central pressure are very small. 

\begin{figure}[ht!]
\begin{center}
\includegraphics[width=10cm]{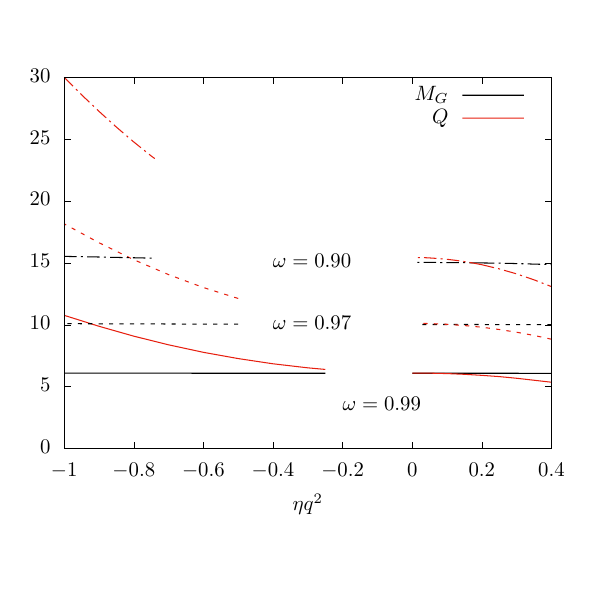}
\end{center}
\vspace{-1.5cm}
\caption{We show the Noether charge $Q$ (red) and  the gravitational mass $M_{G}$ (black),  in dependence of $\eta q^2$ for the boson star solutions 
with $\omega= 0.99$ (solid), $\omega= 0.97$ (dashed) and $\omega=0.90$ (dotted-dashed) respectively. }
\label{fig:m_q_etaq2_2}
\end{figure}

For $\eta q^2 < 0$, we observe the exact opposite: the boson stars can contain many more scalar particles. The Noether charge increases strongly, while the mass remains of the same order of magnitude.  We present some numerical values of our results in Table \ref{table1}.
As can be clearly seen here in combination with the data presented in Fig. \ref{fig:m_q_etaq2_2}, the mass $M_G$ varies only slightly with $\eta q^2$ and  decreases 
when increasing $\eta q^2$. The Noether charge $Q$ on the other hand varies
strongly with $\eta q^2$. Moreover, as can be seen from Fig. \ref{fig:m_q_etaq2_2} a 
gap in $\eta q^2$ exist for which scalarized boson stars are not possible. This gap depends on
the value of the frequency $\omega$ and increases when $\omega$ decreases, i.e.
when $\omega$ decreases.

\begin{figure}[ht!]
\begin{center}
{\includegraphics[width=10cm]{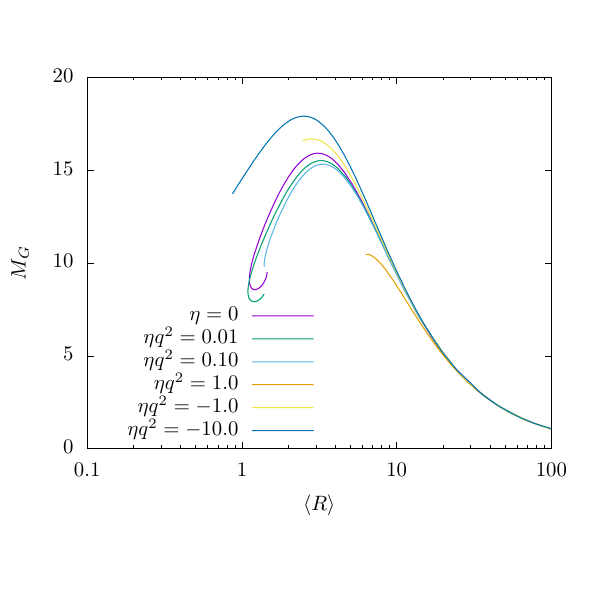}}
\end{center}
\vspace{-1.5cm}
\caption{We show the gravitational mass $M_G$
in function of the mean radius $\langle R\rangle$ of the boson star with time-dependent 
scalar hair for several values of $\eta q^2$. For comparison we also show the
mass-radius relation for the GR limit ($\eta=0$). 
\label{fig:mass_radius}
}
\end{figure}

Finally, and since we want to compare neutron stars with scalar hair with boson stars with scalar hair in this paper, we show the mass-radius relation for the boson stars in Fig. \ref{fig:mass_radius} for several values of $\eta q^2$, see also Table \ref{table1} for
some values.  We find that boson stars with large radius are practically not influenced
by the scalar-tensor coupling, but very compact boson stars are. The radius of the
boson star at maximal mass, $\langle R\rangle^*$ (see Table \ref{table1}) is larger
for all positive $\eta q^2$ that we have studied, however smaller for all negative values of $\eta q^2$. 

If we use the standard argument that a boson star can be thought of as a system of a number $Q$
of scalar particles of mass $m$, we can compare the actual
mass $M_G$ of the boson star and the mass of $Q$ scalar bosons which is $mQ$. 
For $M_G < m Q$, we expect the boson star to form a bound system of these individual
bosons and hence be stable with respect to the decay into those particles. 
Note that with our rescalings, the scalar boson mass $m\equiv 1$. Inspection of Fig. \ref{fig:M_Q_omega} demonstrates that decreasing $\eta q^2$ from zero, the binding between the scalar particles
increases, suggesting that for $\eta < 0$ the non-minimal coupling has effectively an attractive nature.
On the other hand, for $\eta q^2 > 0$, we find that $M_G > Q$ for a part of the second branch of
solutions (see $\eta q^2 = 0.01$) or that -- for sufficiently large $\eta q^2$ -- all boson star solutions are unstable to decay into $Q$ individual bosons (see curves for $\eta q^2 \geq  0.01$.)

\section{Neutron stars}
\label{section:NS}
The energy-momentum tensor for a neutron star is typically assumed to be that of a perfect fluid with $P_r=P_t\equiv P$ and an equation of state (EOS) relating $\rho$ and $P$. 
In addition to the gravity equation (\ref{eq:gravity}), we then also have to solve the Tolman-Oppenheimer-Volkoff (TOV) equation which reads~:
\begin{equation}
           P' = -\frac{b'}{2b}(P + \rho) \  .
\end{equation}
In the following, we will use a polytropic EOS that have already been used previously.
The first EOS, which we will refer to as ``EOSI'' in the following, has been
used in \cite{cdr2015} in the exact same context as in our work and reads~:
\begin{equation}
\label{eq:EOS1}
               \rho = P + K P^{2/3}   \ .
\end{equation}
The second EOS (``EOSII'') has been used in \cite{lemos2004} and is of the form
\begin{equation}
\label{eq:EOS2}
               \rho = \gamma P + K P^{3/5}  \ \ , \ \ \gamma = 0 \ \ {\text or} \ \ \gamma=1\ .
\end{equation}
Note that, although we use the letter $K$ for both EOS, this coupling has different physical dimensions in the two cases. In natural units, the mass dimension of both $\rho$ and $P$ is 
$-2$ and as a consequence, the mass dimension of $K$ is $-2/3$ for the type I and
$-4/5$ for the type II case, respectively.

The radius $R$ of the neutron star is defined differently than that of the boson star. 
Here, the star has a ``hard core'', i.e. a surface outside of which the space-time is given
by the Schwarzschild solution. The relevant conditions to impose in this case are~: 
\begin{equation}
P(R)=0 \ \ , \ \ b(R)=f(R)  \ . 
\end{equation}

To connect the results to physically realistic values for the mass and radius of the neutron stars,
$K$ has to be chosen accordingly. 
However for the purpose of our study, we note that the equations of motion are invariant under the following rescaling~:
\begin{equation}
\label{eq:rescalingNS}
      r \to \lambda r \ \ , \ \ 
			M_G \to \lambda M_G \ \ , \ \ 
			P \to \lambda^{-2} P \ \ , \ \ 
			\rho \to \lambda^{-2} \rho \ \ , \ \ 
			K \to    \lambda^{-\beta} K
\end{equation}
where $\beta=2/3$ for type I and $\beta=4/5$ for type II, respectively. 
Then, a dimensionless  radius $\tilde{R}$ and a dimensionless mass $\tilde{M}_G$ of the configuration 
can be defined according to
\be
\tilde{R} = R K^{1/\beta} \ \ \ , \ \ \ \tilde{M}_G = M_G K^{1/\beta} \ \ .
\ee
Note that we are using natural units here with $\hbar=c=G\equiv 1$.
Reinstalling the natural constants, we find that the mass $M_G$ and $R$ given in Fig. \ref{fig:NS_mass_radius} are related to the dimensionful mass $M_{G,{\rm phys}}$ and
dimensionful radius $R_{\rm phys}$ as follows
\begin{equation}
\frac{M_{G,{\rm phys}} \left[M_{\odot}\right]  } {R_{\rm phys} \left[{\rm  km}\right]}\approx 0.68\frac{M_G}{R}  =  0.68\frac{\tilde{M}_G}{\tilde{R}}\ . 
\end{equation}

\subsection{Numerical results}
In this first part, we will discuss and review already existing results to clarify our construction and compare the two different EOS discussed above in the GR limit.
We will then turn to new scalar-tensor neutron stars using the EOSII for $\gamma=1$. 

\subsubsection{Neutron stars in GR}

\begin{figure}[ht!]
\includegraphics[width=9cm]{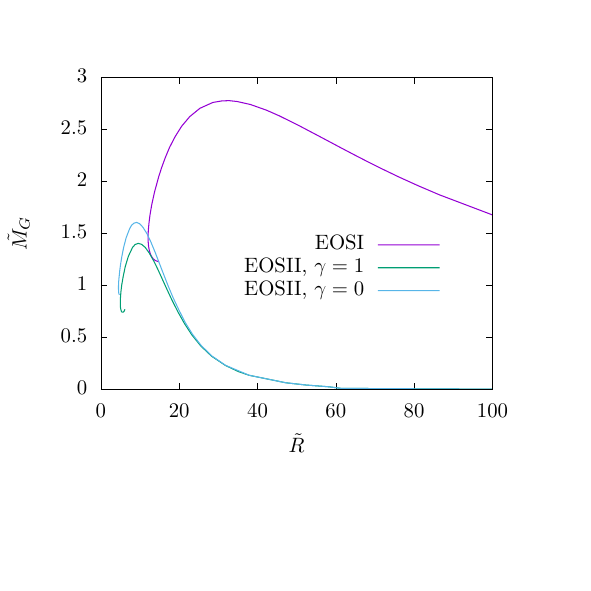}
\includegraphics[width=9cm]{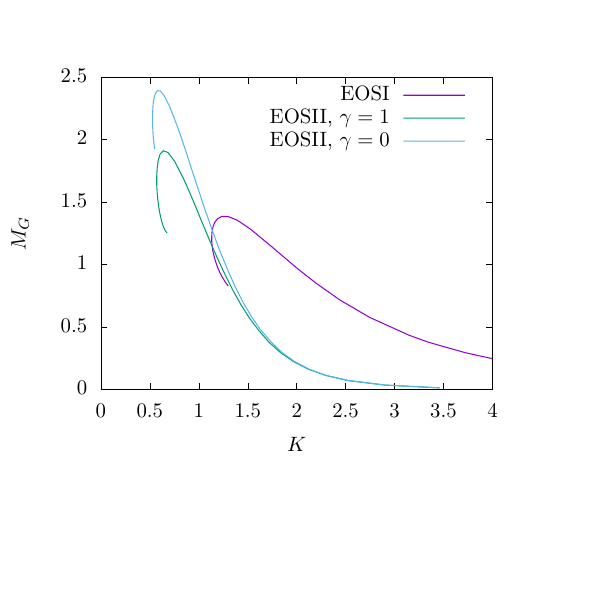}
\vspace{-2.5cm}
\begin{center}
\caption{We show the mass $\tilde{M}_G$ as function of the radius $\tilde{R}$ of the neutron
star solutions for $\eta=0$ and two different EOS, see (\ref{eq:EOS1}) and (\ref{eq:EOS2}), respectively (left). We also show $M_G$ as function of $K$ for the same EOS and
$R=10$ (right). 
\label{fig:NS_mass_EOS}}
\end{center}
\end{figure}

In Fig. \ref{fig:NS_mass_EOS} (left) we show the dimensionless  quantity $\tilde{M}_G$
in function of the radius $\tilde{R}$ of the neutron star in the GR limit and for the two different equations of state. 
Note that using (\ref{eq:rescalingNS}), the axes in this plot have to be rescaled by the same factor $K$ in order to find the physical values of mass and radius of the neutron star.
Contrary to what is presented in \cite{cdr2015}, we find that for a typical neutron
star of radius $R_{\rm phys}=10$km (corresponding to the maximum of the curve) the ratio $M/M_{\odot}\approx 0.6$, and not $M/M_{\odot}\approx 1.2$ as stated in \cite{cdr2015}. Moreover, the qualitative
relation between mass and radius is different to that in 
Fig. 2 of \cite{cdr2015}.

 Comparing e.g. with the gravitational wave detections GW170817 from a binary neutron star merger  \cite{TheLIGOScientific:2017qsa} which
suggests that the two neutron stars in the merger had masses between $0.86 M_{\odot}$ and $2.26  M_{\odot}$
and radii between $10.7$ km and $11.9$ km \cite{Abbott:2018exr} (compare also very new results in \cite{Abbott:2020uma}), we find that the EOS of type I seems to have neutron stars of too low mass. We have hence considered EOSII for
$\gamma=0$, the case studied e.g. in \cite{lemos2004} (albeit for charged neutron stars),
as well as $\gamma=1$, respectively. For $\gamma=0$ our results are in perfect agreement with those of \cite{lemos2004}.
The Fig. \ref{fig:NS_mass_EOS} clearly demonstrates
what for both choices of $\gamma$ we can find neutron stars of radius $10$ km and
mass approx $M_{\odot}$.  Moreover, Fig. \ref{fig:NS_mass_EOS} (right) demonstrates
that for EOSI the value of $K$ should be on the order of $1.5$, while for the EOSII 
it is rather on the order of $0.6$.

\begin{figure}
\begin{center}
\vspace{-1.5cm}
{\includegraphics[width=10cm]{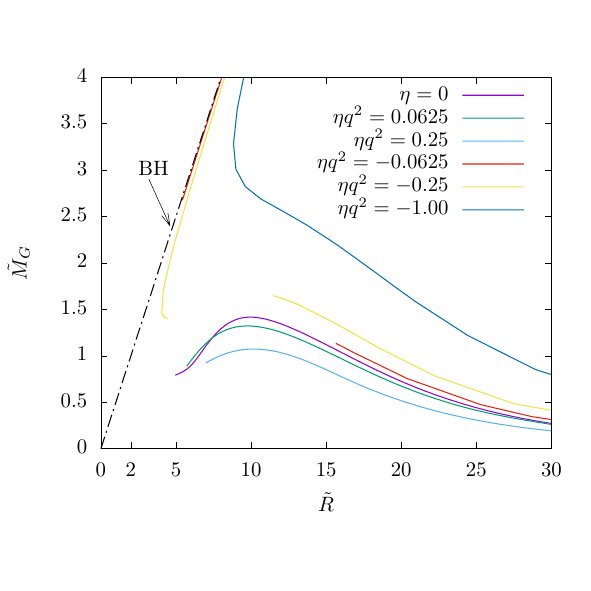}}
\caption{We show the mass $\tilde{M}_G$ as function of the radius $\tilde{R}$ of the neutron
star solutions with EOSII for $\gamma=1$ and different values of $\eta q^2$
including the GR limit ($\eta=0$). The mass-radius relation of the corresponding
Schwarzschild black hole is indicated by ``BH''.
\label{fig:NS_mass_radius}
}
\end{center}
\end{figure}

\begin{figure}[ht!]
\includegraphics[width=9cm]{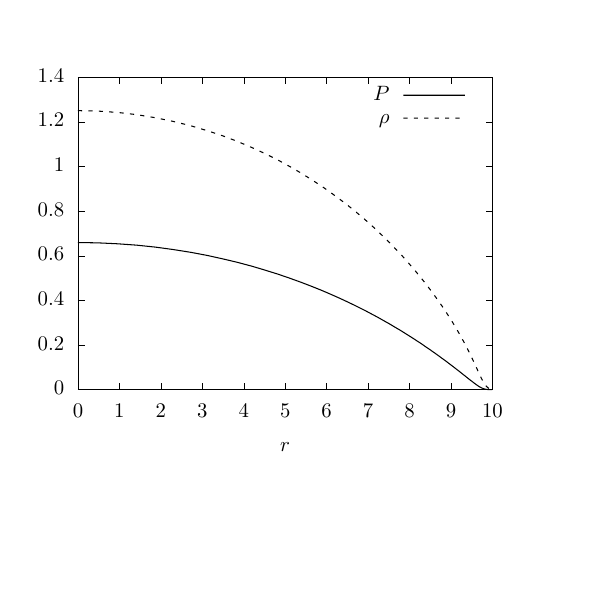}
\includegraphics[width=9cm]{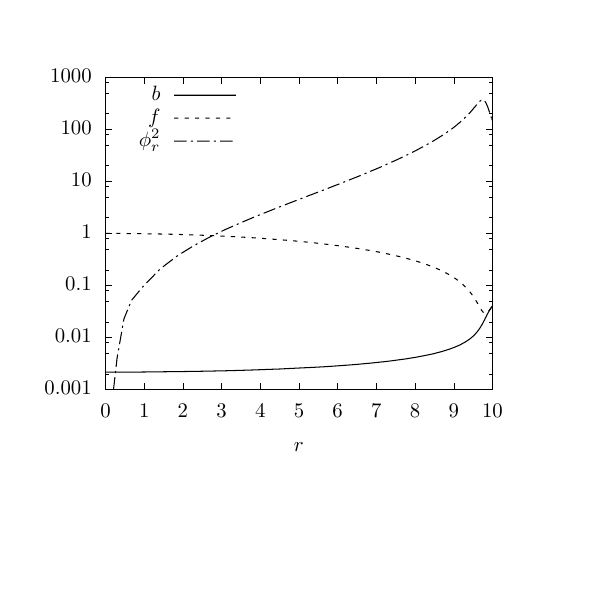}
\vspace{-2.5cm}
\begin{center}
\caption{We show the profiles of the  pressure $P$ (solid) and the energy density $\rho$ (dashed) of a scalar-tensor neutron star with radius $R=10$
for $\eta q^2 =-0.25$,  $P(0) = 0.66$ and type II equation of state (left).  We also show the metric functions 
$b$ (solid) and $f$ (dashed) as well as $\phi_r^2\equiv \phi'^2$ (dotted-dashed) for the same solution (right). 
\label{fig:NS_profiles}}
\end{center}
\end{figure}

\subsubsection{Scalar-tensor neutron stars}
We now turn to the description of the influence of the non-minimal scalar-tensor coupling on the neutron
star solutions constructed with EOSII and  $\gamma=1$.

We find that the existence of neutron stars -- very similar to that of boson stars --
is limited by the requirement of positivity of the denominator in the expansion (\ref{eq:taylor}) for $\eta q^2 > 0$ and by the requirement of positivity of $\phi'^2$ (see (\ref{eq:phip2})) for $\eta q^2 < 0$, respectively. 
Our results for the mass-radius relation of neutron stars for different values of $\eta q^2$ are shown in Fig.\ref{fig:NS_mass_radius}.  The maximal mass $M_{G,{\rm max}}$ of the scalarized neutron stars is reached at roughly the same value of $R\approx 10$, however,
when increasing $\eta q^2$, the value of the maximal mass decreases as compared to the GR limit. 
When decreasing $\eta q^2$ from zero, we find an interesting new phenomenon. Let us choose the
value $\eta q^2 =-0.25$ to explain this in more detail~:
when increasing the central pressure of the star, $P(0)$, we find a branch of solutions 
for $P_0 \leq 0.009$ (in our units) corresponding to $R > 11.4$. 
The solutions constructed for larger $P_0$ (and $R \leq 11.4$) have $(\phi')^2<0$ in some region and are
therefore not acceptable, i.e. we find an interval of $P(0)$ for which
no scalarized neutron stars exist.  Interestingly, we observe that when increasing $P(0)$ sufficiently
(in fact, $P(0) > 0.12$)
a new, second branch of scalarized neutron stars for which $(\phi')^2 > 0$, exists.
The reason for the existence of this new branch can be understood when considering (\ref{eq:phip2}) and the plot of the energy density $\rho$, pressure $P$, the metric functions
$f(r)$ and $b(r)$ as well as $\phi'^2$ given in Fig. \ref{fig:NS_profiles} for neutron star
corresponding to the second branch of solutions. This neutron star has $R=10$ and $P(0)=0.66$. Clearly, all functions are well behaved in particular $\phi'^2 \geq 0$ inside the star.
The reason for the existence of these solutions then also becomes clear~: since
$b(r)$ is very small everywhere inside the star by inspection of (\ref{eq:phip2}) the value of $\phi'^2$ can become positive again. The crucial point is hence the presence of the explicit time-dependence of the 
scalar field, i.e. the fact that $q\neq 0$.
Not surprisingly, these neutron stars are very dense~: as Fig. \ref{fig:NS_mass_radius} demonstrates they are very close to the branch of Schwarzschild black holes. 
When decreasing $\eta q^2$ further, see the curve for $\eta q^2 =-1.0$ in Fig. \ref{fig:NS_mass_radius}, we find that there exists a continuous branch of solutions along which the
central pressure $P(0)$ increases and $(\phi')^2$ stays always positive. Hence, we find
neutron stars that through a continuous deformation of the central pressure can reach mass densities that are very close to that of black holes.

\section{Conclusions}
\label{section:summary}
In this paper, we have studied the properties of boson and neutron stars in a scalar-tensor gravity models which contains an explicitly time-dependent real scalar field. The norm of the Noether current
associated to the shift symmetry of the gravity scalar is finite everywhere in the space-time.
We find that the explicit time-dependence does allow non-trivial scalar fields to exist in both the space-time of a boson star and neutron star, respectively. Moreover, the presence of the gravity scalar has interesting consequence for the properties of these objects. While the boson star's
mass does not vary strongly when increasing or decreasing the scalar-tensor coupling from zero, it has a large effect on the number of scalar bosonic particles making up the boson star, the mean radius and central density and pressure. This means that while in the GR limit, boson stars of the type studied here, so-called ``mini boson stars'', have radius of a few Schwarzschild radii (see e.g. \cite{bosonstars}), the
radius of the scalar-tensor counterparts could, in fact, be much closer to the Schwarzschild radius. 

For neutron stars, we have investigated a specific equation of state that allows for neutron stars with mass around $10$km and a few solar masses, an assumption that seems to be realistic
from recent gravitational wave detections \cite{TheLIGOScientific:2017qsa,Abbott:2018exr, Abbott:2020uma}. While neutron stars have a ``hard core'' outside which the pressure is
strictly zero, the change of properties is comparable to that of boson stars.
In particular, for negative scalar-tensor coupling and the gravity scalar changing slowly in time, we find that new branches of solutions of neutron stars exist that have a mass-radius relation very close
to that of Schwarzschild black holes. Increasing the time change of the gravity scalar, we find
that we can continuously deform ``standard'' mass neutron stars to these objects with
large central pressure $P(0)$.

In summary, our results indicate that the presence of a gravity scalar in the case of globally regular, compact objects prevents these objects from collapsing to a black hole at the values
known in GR due to an increased central pressure allowed inside the stars.

\vspace{0.5cm}
{\bf Acknowledgments} 
BH would like to thank FAPESP for financial support under grant {\it 2019/01511-5}.


\end{document}